\documentstyle[psfig,draft]{mn}

\def\gs{\mathrel{\raise0.35ex\hbox{$\scriptstyle >$}\kern-0.6em
\lower0.40ex\hbox{{$\scriptstyle \sim$}}}}
\def\ls{\mathrel{\raise0.35ex\hbox{$\scriptstyle <$}\kern-0.6em
\lower0.40ex\hbox{{$\scriptstyle \sim$}}}}

\title[The diversity of submm galaxies]
      {The diversity of SCUBA-selected galaxies}

\author[R.\,J.\ Ivison et al.]
       {R.\,J.\ Ivison,$^{\! 1}$ Ian Smail,$^{\! 2}$ A.\,J.\ Barger,$^{\! 3}$ 
	J.-P.\ Kneib,$^{\! 4}$  A.\,W.\ Blain,$^{\!5}$ 
        F.\,N.\ Owen,$^{\! 6}$ \and 
        T.\,H.\ Kerr$^{\! 7}$ and L.\,L.\ Cowie$^{3}$
        \vspace*{1mm}\\
        $^1$ Department of Physics \& Astronomy, University College London, 
	Gower Street, London WC1E 6BT\\
        $^2$ Department of Physics, University of Durham, South Road,
        Durham DH1 3LE\\
	$^3$ Institute for Astronomy, University of Hawaii, 2680 Woodlawn
        Drive, Honolulu, HI 96822, USA\\
        $^4$ Observatoire de Toulouse, 14 avenue E.\ Belin,
        31400 Toulouse, France\\
        $^5$ Institute of Astronomy, Madingley Road, Cambridge CB3 0HA\\
        $^6$ National Radio Astronomy Observatories, P.O.\ Box O, Socorro,
        NM 87801-0387, USA\\
        $^7$ Joint Astronomy Centre, 660 N.\ A`oh\={o}k\={u} Place,
        University Park, Hilo HI\,96720, USA}

\date{Accepted ... ; Received ... ; in original form 1999 June 1}

\pagerange{000--000}

\begin{document}

\maketitle

\begin{abstract}
We present extensive observations of a sample of distant,
submillimetre (submm) galaxies detected in the field of the massive
cluster lens, Abell\,1835, using the Submm Common-User Bolometer Array
(SCUBA). Taken in conjunction with earlier observations of other
submm-selected sources (Ivison et al.\ 1998; Smail et al.\ 1999;
Soucail et al.\ 1999) we now have detailed, multi-wavelength
observations of seven examples of the submm population, having
exploited the combination of achromatic amplification by cluster
lenses and lavish archival datasets. These sources, all clearly at
$z\gs1$, illustrate the wide range in the radio and optical properties
of distant submm-selected galaxies. We include detailed observations
of the first candidate `pure' starburst submm galaxy at high redshift,
a $z=2.56$ interacting galaxy which shows no obvious sign of hosting
an AGN. The remaining sources have varying degrees of inferred AGN
activity (three from seven of the most luminous show some evidence of
the presence of an AGN) although even when an AGN is obviously
present it is still not apparent if reprocessed radiation from this
source dominates the submm emission.  In contrast with the variation
in the spectral properties, we see relatively homogeneous morphologies
for the population, with a large fraction of merging or interacting
systems. Our study shows that virtually identical spectral energy
distributions are seen for galaxies which exhibit strikingly different
optical/UV spectral-line characteristics. We conclude that standard
optical/UV spectral classifications are misleading when applied to
distant, highly obscured galaxies and that we must seek other means of
determining the various contributions to the overall energy budget of
submm galaxies and hence to the far-infrared extragalactic background.
%A picture of the submm galaxy population is emerging where virtually
%identical spectral energy distributions (SEDs) are seen for galaxies
%with very different optical/UV spectral-line characteristics,
%indicating that classification on the basis of easily obscured line
%emission may be misleading and that we must seek other means of
%determining the various contributions to the overall energy budget.
\end{abstract}


\section{Introduction}

Following the commissioning of SCUBA (Holland et al.\ 1999) on the
James Clerk Maxwell Telescope (JCMT) in 1996--97, surveys of the
distant Universe have rapidly increased the sample of submm-selected
galaxies. The survey on which this paper is based employed massive,
intermediate-redshift clusters to magnify submm sources in the distant
Universe and yielded a sample of seventeen strongly star-forming,
dusty galaxies, mostly at $z>1$ (Smail, Ivison \& Blain 1997; Smail et
al.\ 1998a; Barger et al.\ 1999a; Blain et al.\ 1999a,b). This
population is responsible for the bulk of the mm/submm background
detected by {\em COBE} (Puget et al.\ 1996; Fixsen et al.\ 1998) as
discussed by Blain et al.\ (1999a).  Subsequent blank-field surveys
have achieved comparable source-plane sensitivities (Barger et al.\
1998, 1999b; Hughes et al.\ 1998; Eales et al.\ 1999) and have roughly
tripled the number of submm-selected galaxies. As a result of these
efforts, a broad consensus has already been reached on the surface
density of submm galaxies down to the practical confusion limit for
detection of $\sim$1.5\,mJy (Blain, Ivison \& Smail 1998).

There are several advantages to mapping the distant submm Universe
through massive, well-constrained cluster lenses.  The most relevant
for this study is that the achromatic nature of the gravitational
amplification means that not only is the effective depth of the submm
maps increased, but that the counterparts at all other wavelengths are
similarly amplified, allowing detailed follow-up observations to be
obtained using the current generation of telescopes and
instrumentation.  A broad variety of follow-up observations are
necessary to correctly categorise the submm sources. By obtaining
redshifts, constructing complete SEDs, and using other diagnostics,
the evolutionary status of the galaxies can be explored and the
dominant source of power can be ascertained, be it gravitational (AGN)
or stellar (starburst).

The brightest of the sources found so far with SCUBA was also the
first to be discovered --- SMM\,J02399$-$0136 (Ivison et al.\ 1998 ---
I98), a $z=2.8$ interacting galaxy behind the cluster Abell\,370 with
an intrinsic luminosity in excess of $10^{13}$\,L$_{\odot}$. An
amplification factor of 2.4 (roughly a magnitude) brought SMM\,J02399
within reach of spectrographs on 4-m telescopes, and the presence of a
dust-enshrouded AGN was immediately inferred from the high-excitation
emission lines which dominate its rest-frame UV spectrum.  This
classification was subsequently confirmed by the characteristics of
the radio counterparts seen in deep 5- and 1.4-GHz maps (Ivison et
al.\ 1999; Owen et al., in prep). Despite the presence of an AGN,
there is also evidence that significant star-formation activity has
taken place and may be continuing. The presence of a large quantity of
dust points to past activity, while the strength of the Balmer
H$\alpha$ emission line suggests a current star-formation rate (SFR)
of order 10$^3$\,M$_{\odot}$\,yr$^{-1}$.  The detection of a massive
gas reservoir in this galaxy (Frayer et al.\ 1998) confirms that such
a high SFR could be sustained for a considerable period of time, $\gs
10^8$\,yrs.

Apart from obscured AGN, the only other components of the submm
population so far identified by SCUBA surveys are a class of
optically-faint but infrared-bright galaxies with extremely red
colours (those with $I-K>6$, e.g.\ Elston, Rieke \& Rieke 1988).
Brighter than $K\sim 20.5$, these galaxies could comprise 10 per cent
of the submm population down to fluxes of $\sim 1$\,mJy at 850\,$\mu$m
(Smail et al.\ 1999).  The broad spectral properties of this class are
consistent with them being the high-redshift analogues of local
ultraluminous galaxies (Dey et al.\ 1999; Smail et al.\ 1999).
Moreover, they may account for 5 per cent or more of the total
extragalactic background at all wavelengths, a small but
non-negligible fraction.

%
% FIGURE 1
%
\begin{figure*}
\centerline{\psfig{file=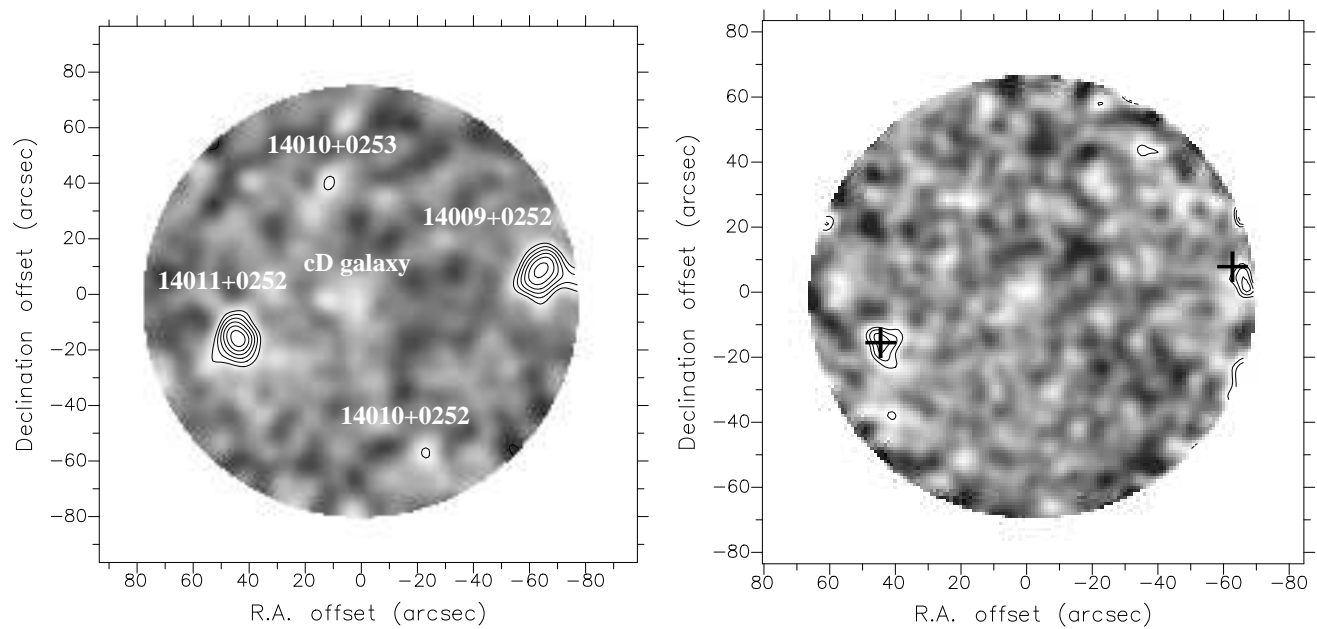,angle=0,width=7in} }
\caption{The central region of the Abell\,1835 SCUBA field at
850\,$\mu$m ({\em left}) and 450\,$\mu$m ({\em right}). The submm
sources discussed in the text are labelled on the 850-$\mu$m
image. The 450-$\mu$m image has been smoothed with a 4.5$''$ Gaussian
to give a 9$''$ FWHM effective resolution; the brightest 850-$\mu$m
source centroids are marked.  Emission associated with the cD galaxy
in the centre of Abell\,1835 is discussed by Edge et al.\ (1999). The
faint 850-$\mu$m sources to the north and south did not meet the
selection criteria of the most liberal Smail et al.\ (1998a) sample
(50-per-cent completeness, nominally 3\,$\sigma$ in a 30$''$
aperture). We discuss these sources further in \S3.  Contours are
plotted at: (a) $-3, 3, 4, 5, 6, 7$ and $8\sigma$, where $\sigma =
1.7$\,mJy\,beam$^{-1}$, and (b) $-3, 3, 4$ and $5\sigma$, $\sigma =
6.0$\,mJy\,beam$^{-1}$. The field centres are at R.A.\ $14^{\rm h}
01^{\rm m} 02.^{\rm s}02$, Dec.\ $+02^{\circ} 52' 41.''6$ (J2000).}
\end{figure*}

In this paper, we describe detailed follow-up observations of a
sub-sample of submm-selected galaxies discovered in the complete
survey of Smail et al.\ (1998a) which comprised deep submm maps of
seven massive clusters, reaching median 1\,$\sigma$ sensitivities
around 0.7\,mJy in the source planes. These particular galaxies lie
behind the luminous X-ray cluster, Abell\,1835, at $z=0.25$. Without
the amplification due to the cluster, their redshift identification
and follow-up would have been challenging even for 10-m
telescopes. Aided by the lensing, however, we are able to investigate
their optical, infrared (IR) and radio morphologies, putting together
spectral energy distributions (SEDs) spanning the rest frame from the
Lyman limit to the radio.
 
The layout of this paper is as follows. First we describe optical, IR,
submm and radio measurements of the submm sources (\S2) and present
the identification of their optical/IR/radio counterparts in \S3.  In
\S4 we then discuss their properties in the context of other luminous,
dusty galaxies, and speculate on the likely source of power for their
prodigious far-IR luminosities.  Finally, we discuss the emerging
picture of the submm galaxy population and how we can hope to refine
our knowledge of these important contributors to the evolution of the
co-moving luminosity density of galaxies (\S5). We assume $q_0 = 0.5$
and $H_0 = 50$\,km\,s$^{-1}$\,Mpc$^{-1}$ throughout.

\section{Observations}

%
% FIGURE 2
%
\begin{figure*}
\centerline{\psfig{file=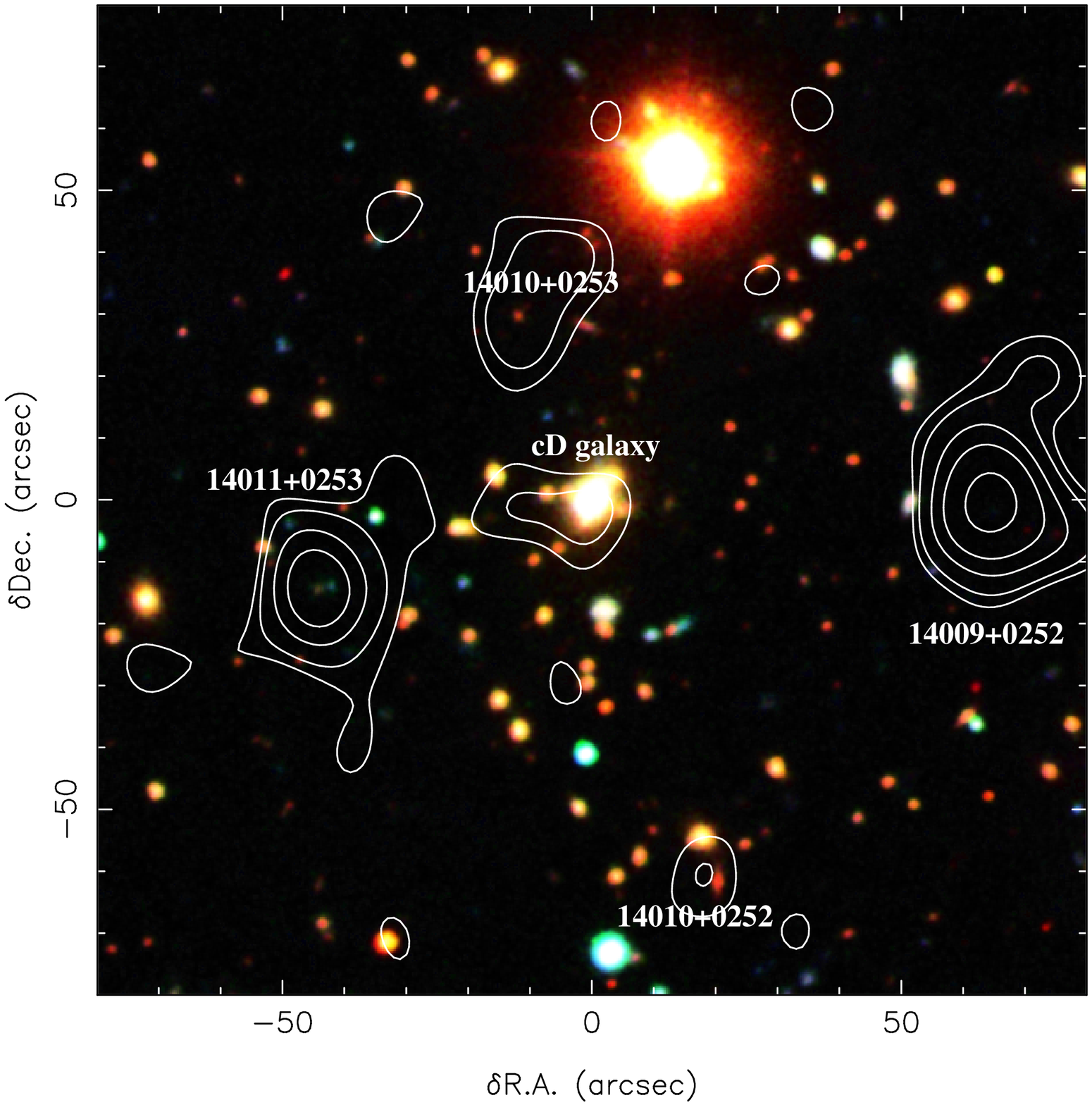,width=6.0in,angle=0}}
%{file=figs/a1835i_850_newS3b.cps,width=6.8in,angle=0}}
\caption{A true-colour image of Abell\,1835 constructed from $UBI$
images taken on the P200 telescope (\S2.2). The 850-$\mu$m SCUBA map,
smoothed to $20''$ FWHM, is overlayed as a contour plot. There are
possible optical counterparts to at least four of the submm sources,
with the brightest submm source having none.  Note also that the
slight offset in the position of the central submm-bright cluster
galaxy is caused by reference signals from the two brightest sources,
which overlapped $\sim10''$ west of the central cluster galaxy
whenever the field was observed close to transit.  The central galaxy
is at R.A.\ $14^{\rm h} 01^{\rm m} 02.^{\rm s}11$, Dec.\ $+02^{\circ}
52' 43.''1$ (J2000).}
\end{figure*}

\subsection{SCUBA and Very Large Array observations}

A 23-ks jiggle-map integration of Abell\,1835 was taken in outstanding
conditions during 1998 January and April with SCUBA on the JCMT,
recording data at both 450 and 850\,$\mu$m. The 850-$\mu$m zenith
atmospheric opacity, measured each hour with a skydip, was typically
0.15 and never worse than 0.20. The pointing was checked regularly
using the blazar, 1413+135, and was very stable. Calibration maps of
Uranus and the unresolved secondary calibrator, CRL~618, were also
obtained.

The data were reduced using the {\sc surf} software package (Jenness
\& Lightfoot 1998). The rms map sensitivity at 850\,$\mu$m was
1.7\,mJy\,beam$^{-1}$ and two bright sources were detected (Fig.~1) at
a significance of $\sim8\sigma$. These sources, SMM\,J14009+0252 and
SMM\,J14011+0252 (hereafter SMM\,J14009 and SMM\,J14011), are
discussed later (\S3). A conservative estimate of the 1-$\sigma$
positional uncertainty is $3''$, a value which includes the combined
statistical and systematic errors. This is commensurate with the
astrometric shifts seen in deep SCUBA images of known submm sources,
e.g.\ HR\,10, 4C\,41.17, 8C\,1435+635 (Ivison et al., in prep) and
HDF850.1 (Downes et al.\ 1999).

SMM\,J14011 was also detected in the 450-$\mu$m map (Fig.~1).  When
smoothed slightly (to 9$''$ FWHM), the noise level at 450\,$\mu$m is
6.0\,mJy\,beam$^{-1}$ and the significance of the SMM\,J14011 detection
is above $5\sigma$ (above $7\sigma$ using an aperture-based flux
estimate, which is the most reliable method for faint sources at
450\,$\mu$m). The source is coincident with the 850-$\mu$m position to
within the statistical uncertainty of the centroiding on the 850-$\mu$m
beam.

Although the 450- and 850-$\mu$m maps were obtained simultaneously,
the field of view at 450\,$\mu$m is slightly smaller than that at
850\,$\mu$m (see Fig.~1); as a result, SMM\,J14009 lies close to the
edge of the 450-$\mu$m map. There is a 5-$\sigma$ 450-$\mu$m source
($\sim32$\,mJy) $6''$ south of the 850-$\mu$m position ($4''$ south of
the 1.4-GHz position --- see later). We view it as likely that this
emission is due to SMM\,J14009: the brightest peak at 450\,$\mu$m
undoubtedly exagerates the positional offset; if the 450-$\mu$m
position was taken as the centre of the 3-$\sigma$ contour the offset
would not be significant.

More observations of these two bright sources were made in 1998
February and 1999 February using the photometry mode of SCUBA. As
described by I98, the secondary mirror was chopped, at 7.8\,Hz in this
instance, using a simple `filled-square' 9-point jiggle with $2''$
offsets in both the signal and reference beams, with the telescope
nodding between the two every 9\,s in a
signal--reference--reference--signal pattern. The total integration
times were: SMM\,J14011, 1350\,$\mu$m, 5.4\,ks; SMM\,J14009,
1350\,$\mu$m, 2.7\,ks; SMM\,J14009, 450 and 850\,$\mu$m, 2.7\,ks. The
photometry-mode flux densities agree with those determined from the
450- and 850-$\mu$m maps.

A third submm source was detected in the 850-$\mu$m maps close to the
bright central cluster galaxy in Abell\,1835. The offset from the
optical position (Fig.~2) was caused by reference signals from the two
brightest sources, which overlapped $\sim10''$ west of the central
cluster galaxy whenever the field was observed close to transit, i.e.\
for a significant fraction of the total integration time.  The
properties of this galaxy, and the submm-detected central galaxy in
the cluster Abell\,2390, are presented by Edge et al.\ (1999). Since
we are concerned solely with the field submm galaxy population, we
shall not discuss this object again, save to identify it on our
wide-field optical, radio and submm images.

Two other faint submm sources are just visible in Figs~1 \& 2, to the
north (SMM\,J14010+0253) and south (SMM\,J14010+0252) of the cluster
centre.  Both of these sources are fainter than the flux limit of
$S_{850}\geq 5.0$\,mJy adopted for the Abell\,1835 field in Smail et
al.\ (1998a), nominally associated with the 50-per-cent completeness
limit. However, as we discuss below there is reason to believe that
both sources are reliable detections and that the selection used by
Smail et al.\ (1998a) might be more conservative than claimed.  This
would result from a combination of the relatively large detection
aperture, 30$''$ diameter or 2 beam widths, used to construct the
original 850-$\mu$m catalogues (which will reduce the apparent
significance of faint point sources) and the variation in the sky
noise across the submm maps (the object selection algorithm adopted
the rms in the noisiest areas of the map).

Abell\,1835 was observed at 1.4\,GHz with the National Radio Astronomy
Observatory's (NRAO) Very Large Array (VLA) in B configuration during
1998 April.  These observations were part of a survey of all seven
clusters in the SCUBA sample and were specifically intended to provide
radio detections and accurate positions for any SCUBA sources within
the field (Smail et al.\ 2000). To this end, a very deep 29-ks
integration was obtained, a factor two from the deepest 1.4-GHz map
ever obtained (Richards 2000). The phase calibrator was 1354$-$021,
for which the position is known to within $0.002''$, and the data were
calibrated in flux using 1331+305 (3C\,286).

In order to reach sufficiently low noise levels, the process of data
acquisition, reduction, self calibration, mapping and cleaning is
complex, involving the mapping of several dozen sources throughout the
primary beam. This process is discussed in detail by Morrison
(1999). The resulting 1.4-GHz map of Abell\,1835 (Fig.~3) has a relatively
uniform noise level of 16\,$\mu$Jy\,beam$^{-1}$, with a synthesized
beam measuring $5.1'' \times 4.5''$ FWHM at a position angle (PA) of
$120^{\circ}$.

A close comparison of the submm maps with the deep radio map shows
that four of the five submm sources have associated radio emission
(Fig.~3).  The high fraction of radio counterparts to faint submm
sources underlines the usefulness of very deep radio maps for
identification of submm galaxies (I98; Smail et al.\ 2000).  Ordered
in terms of their radio flux, the detected submm sources are the
central cluster galaxy; the weak submm source SMM\,J14010+0252 (with
an unresolved radio counterpart $3.8''$ north of the nominal submm
position) and the two bright submm sources SMM\,J14009 and
SMM\,J14011, both of which have radio counterparts within 1--2$''$.
The very faint submm source, SMM\,J14010+0253, lacks a radio
counterpart but an obvious optical counterpart can be seen in Fig.~2,
a pair of faint, red galaxies separated by $\sim2''$ (see \S3.3).

%
% FIGURE 3
%
\begin{figure}
\centerline{\psfig{file=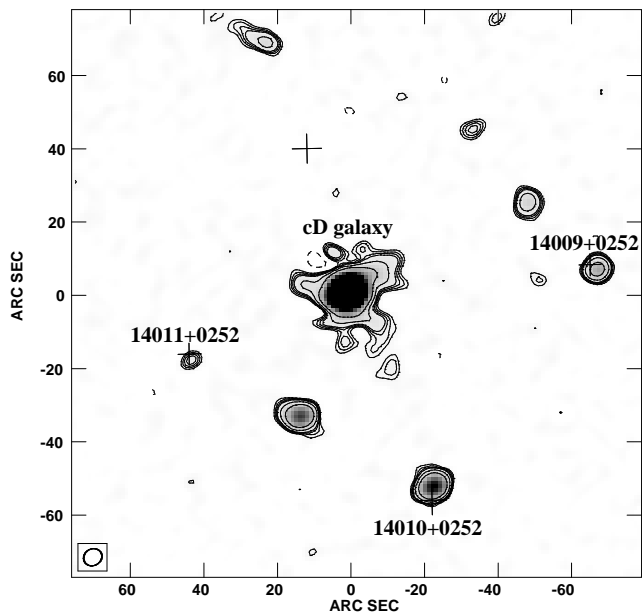,width=3.4in,angle=0}}
\caption{The central region of the Abell\,1835 field at 1.4\,GHz.  The
850-$\mu$m positions of SMM\,J14009 and SMM\,J14011 are shown by
crosses.  The central galaxy is associated with weak submm emission,
while the strong radio source to the south appears to coincide with
the faint submm source SMM\,J14010+0252. Contours are plotted at $-3,
3, 4, 5, 6, 10$ and $20\sigma$, where $\sigma =
16$\,$\mu$Jy\,beam$^{-1}$. The field centre is at R.A.\ $14^{\rm h}
01^{\rm m} 02.^{\rm s}02$, Dec.\ $+02^{\circ} 52' 41.''6$ (J2000).}
\end{figure}

\subsection{Optical and near-IR imaging}

Abell\,1835 is one of the most luminous X-ray clusters known, with
$L_X (2$--$10\,{\rm keV}) \sim 4.5 \times 10^{45}$\,ergs\,s$^{-1}$
(Allen 1998) and it harbours a massive cooling flow. An archival
2.8-ks {\it ROSAT} {\it HRI} X-ray image of the cluster exists and we
use this to place limits on the X-ray emission from the various submm
sources (\S4.2).  Due to its high X-ray luminosity and hence high
central mass, the cluster was included in an optical imaging survey
for strongly lensed features (Smail et al.\ 1998b; Edge et al., in
prep) and we exploit those images here.
 
The data consist of $UBI$ images taken with the COSMIC imaging
spectrograph (Kells et al.\ 1998) on the 5.1-m Hale Telescope (P200)
at Palomar Observatory; total integration times were 3.0\,ks in $U$,
0.5\,ks in $B$ and 1.0\,ks in $I$, the latter with $1.1''$ seeing
(Fig.~2).  More information is given in Smail et al.\ (1998b).

Additional optical imaging was acquired on the night of 1998 July 27
at the 4.2-m William Herschel Telescope on La Palma.  We used a
1\,k$^2$ TEK CCD on the auxilary Cassegrain port giving a pixel scale
of $0.11''$ pixel$^{-1}$.  A total of three 300-s exposures were
obtained in $R$ under good conditions.  These frames were reduced in a
standard manner, flatfielded with twilight flats, aligned and
combined. The seeing measured off the final stacked frame is $0.68''$
FWHM and this therefore provides a relatively clear view of the
rest-frame optical/UV morphology of any distant submm galaxies in the
field.

Near-IR observations of this field were obtained to determine the
rest-frame optical morphologies and luminosities of the submm
galaxies.  These observations were undertaken typically under good
conditions during several observing runs in 1998 July and 1999
February--April on the 3.8-m UK Infrared Telescope (UKIRT), Mauna Kea,
using the IRCAM3 and UFTI cameras (Leggett 1998). The observations
used dithered sequences of images (on a 10$''$ non-repetitive grid) to
allow the science images to be flatfielded using a running median sky
frame constructed from the images themselves.  The frames were linearised, 
dark subtracted, flatfielded and combined in a standard manner.  The final
stacked images were photometrically calibrated using UKIRT Faint
Standards (Casali \& Hawarden 1992).  More details of the observations
are given in Table~1.

%
% Table 1
%
\begin{table}
\begin{center}
\caption{\hfil Log of UKIRT imaging observations. \hfil}
\begin{tabular}{lccccc}
\noalign{\medskip}
Target & Instrument & Date & Band &  t$_{\rm exp}$ & Seeing \cr
& & & & (ks) & ($''$) \cr
\noalign{\medskip}
J1/J2 & IRCAM3 & 98 Jul 11 & $J$ & 3.2 & $\sim1.5$ \cr
J1/J2 & IRCAM3 & 99 Feb 10 & $K$ & 7.0 & 0.60 \cr
J3/J4 & IRCAM3 & 99 Feb 08 & $K$ & 9.2 & 0.75 \cr
J6/J7 & UFTI   & 99 Feb 28 & $K$ & 2.4 & 0.50 \cr
J8    & UFTI   & 99 Mar 02 & $K$ & 2.4 & 0.45 \cr
\noalign{\smallskip}
\end{tabular}
\end{center}
\end{table}

\subsection{Optical and near-IR spectroscopy}

Spectroscopy of the optical counterpart of SMM\,J14011, J1/J2 (see
\S3.1), and SMM\,J14010+0253, J6/J7 (see \S3.3), was undertaken with
the Low-Resolution Imaging Spectrometer (LRIS) multi-object
spectrograph (Oke et al.\ 1995) on the 10-m Keck-{\sc ii} telescope,
Mauna Kea, during 1998 July 18 and 1999 August 4, as part of the
spectroscopic survey of candidate optical counterparts to the submm
sources in our sample (Barger et al.\ 1999).  A 300-l\,mm$^{-1}$
grating blazed at 5000\,\AA\ and a $1.5''$-wide long-slit were used,
yielding an effective resolution of about 14\,\AA\ and a total
wavelength range of 3800--8700\,\AA.  For SMM\,J14011, the slit was
placed along the line joining J1/J2 and J3 (\S3.2), at a PA of
$102^{\circ}$, and the total exposure time was 3.6\,ks. For
SMM\,J14010+0253, the slit was placed along the line joining J6/J7, at
a PA of $145^{\circ}$, and the total exposure time was 1.8\,ks. The
objects were stepped along the slit by 10\,arcsec in each direction,
and the sky backgrounds were removed using the median of the images to
avoid the problems of flat-fielding LRIS data. One-dimensional spectra
were optimally extracted. Flux calibration of the spectra was achieved
using observations of Oke (1990) standards. Details of the
spectroscopic reduction procedures can be found in Cowie et al.\
(1996).

Near-IR spectra of SMM\,J14011 and SMM\,J14010+0253 were obtained
with CGS4  on UKIRT during 1998 August 03, 1999
February 09--12 and 1999 April 07--09 during good or moderate
conditions. The total exposure time in $K$ for SMM\,J14011 was 2.9\,ks
during 1998 August 03, 8.2\,ks during 1999 February 09 and 19.7\,ks
during 1999 February 11--12. For SMM\,J14010+0253, 15.4 and 15.1\,ks of
exposure time were obtained during 1999 April 07--08 in the $K$ and $H$
bands, respectively.

The 40-lines\,mm$^{-1}$ grating was used to cover the entire $K$ or $H$
windows.  A $90''$-long $1.22''$-wide slit was used, orientated to
cover both optical components of each source (i.e.\ PAs of
$110.0^{\circ}$ for SMM\,J14011 and $137.3^{\circ}$ for
SMM\,J14010+0253). During 1999 February 09 and 11--12, a PA of
$12.5^{\circ}$ was used for SMM\,J14011 to cover the major axis of its
dominant near-IR component. The data were Nyquist sampled and the
telescope was nodded $18.3''$ along the slit every 60 or 80\,s to allow
adequate sky subtraction. Offsets from nearby stars were used to
position the slit on the galaxies. Flux calibration and telluric line
cancellation were performed by ratioing the spectrum with that of
HD\,129655 or BS\,5275 then multiplying by blackbody spectra
appropriate to the temperature and magnitude of those stars. Wavelength
calibration is accurate to much better than 0.001\,$\mu$m.

%
% FIGURE 4
%
\begin{figure*}
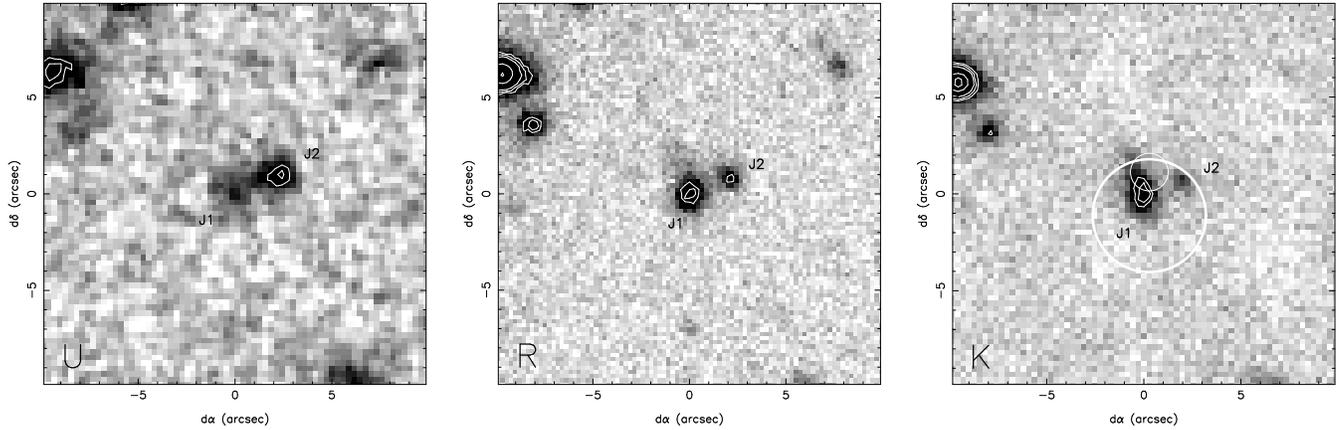

\centerline{
\psfig{file=fig4a.eps,width=2.2in,angle=270} \hspace*{0.1in}
\psfig{file=fig4b.eps,width=2.2in,angle=270} \hspace*{0.1in}
\psfig{file=fig4c.eps,width=2.2in,angle=270}}
\caption{$U$, $R$ and $K$ images of SMM\,J14011 with the two
optical/near-IR counterparts, J1 and J2, identified.  Note that J2
appears relatively compact, even in the $\sim0.6''$ seeing of the $R$
and $K$ exposures. The low-surface-brightness extension to the North
of J1 may be either a further companion or the remnant of a tidal
interaction between J1 and J2. It is prominent in $K$ which suggests a
possible emission-line contribution from redshifted Balmer
H$\alpha$. In the $K$ image, the large circle shows the positional
uncertainty for the submm source; the small circle (radius 1$''$)
shows the combined radio and IR positional uncertainty. The
panels are $20''$ square; north is up, east to the left. The $U$-band
image has been smoothed with a 0.35$''$ FWHM Gaussian.}
\end{figure*}

\section{Results}

We now present the identification of the counterparts of the distant
submm galaxies at optical, IR and radio wavelengths.  We list the
positions and fluxes of the counterparts in the radio, submm, far- and
near-IR, optical and X-ray bands in Table~2.  For optical and
near-IR counterparts, Table~2 gives photometry in $3''$-diameter
apertures from the seeing-matched $U\!BRI\!JK$ images (where
available). The reddening towards Abell\,1835 is $E(B-V)=0.05$, and no
foreground reddening correction has been applied to the results in Table~2.
All values are uncorrected for lens amplification (see \S4).

\subsection{SMM\,J14011+0252 (J1/J2)}

The 1.4-GHz image shown in Fig.~3 clearly shows an unresolved faint
source (FWHM\,$< 2.9''$) within $2''$ of the submm position of
SMM\,J14011. The peak surface brightness is $115 \pm
16$\,$\mu$Jy\,beam$^{-1}$ and we estimate the total uncertainty in the
integrated flux density to be 30\,$\mu$Jy.  Using the accurate
astrometry from our 1.4-GHz map and our deep Palomar and WHT images
(\S2.2) we identify two faint optical counterparts to SMM\,J14011,  J1
and J2 (Fig.~4). In our $I$-band image, J1 is only $1.1''$ from the
nominal radio position, with J1 and J2 separated by $2.1''$ at a
position angle of $\sim$110$^{\circ}$.

The continuum slope in the rest-frame mid-UV (longward of the Lyman
limit) for J1 is $\alpha \sim -1.3$ ($\alpha$, where $S_{\nu} \propto
\nu^{+\alpha}$) while J2 shows $\alpha \sim -0.4$. Using the near-IR
broadband photometry (Table~2) we can also estimate the continuum
slope into the rest-frame optical where we find spectral indices of
$-1.6$ and $-1.0$ for J1 and J2, virtually identical to those seen for
the components of SMM\,J02399, and significantly redder than typically
seen in Lyman-break galaxies ($\alpha \sim -0.3$).  The $K$-corrected
distance modulus at $z=2.56$ is 47.34 (for a spectral index of $-1.6$)
and so J1 and J2 have a combined absolute magnitude of $M_R = -25.0$
or $M_K=-28.3$.

In passbands longward of the $U$ band, J1 is brighter than J2 (Fig.~4;
Table~2) and well resolved, with an intrinsic FWHM of $1.05 \pm
0.05''$. The fainter counterpart, J2, shows a more compact morphology,
although it too appears to be marginally resolved with a deconvolved FWHM of
$0.4 \pm 0.1''$.  J1 exhibits a low-surface-brightness extension
stretching $3''$ to the north, which appears to be somewhat more
prominent in $K$ than $R$ (Fig.~4).

The LRIS spectrum of J1 and J2 shows a number of narrow emission lines
and absorption features superimposed on a strong continuum (Fig.~5;
Table~3).  Using all the features listed in Table~3 we estimate
redshifts of $z=2.56$ for both J1 and J2\footnote{This redshift is
consistent with limits of $1.7\ls z\ls 3.8$ and $2.2<z<6.3$ derived
respectively from the observed radio--submm spectral index of
$\alpha^{850}_{1.4} =0.85\pm 0.05$ and the 450- to 850-$\mu$m flux
ratio, $2.9\pm 0.9$.} and conclude that they are physically
associated.  The emission lines are unresolved, with intrinsic line
widths of $<300$\,km\,s$^{-1}$ in the rest frame. We see no evidence
for a broad component of these lines.  The continuum shape of J1 and
J2 is consistent with the broad-band photometry, with J1 having a
substantially redder UV spectral slope (Table~2 and Fig.~5).

We also see a single narrow line, superimposed on a strong continuum,
in the $K$-band spectrum of J1/J2 from UKIRT (Fig.~6). The line is
unresolved ($<$290\,km\,s$^{-1}$ FWHM after deconvolving the
560-km\,s$^{-1}$ instrumental resolution) and its wavelength, $2.340
\pm 0.001$\,$\mu$m, is consistent with Balmer H$\alpha$ at $z=2.565
\pm 0.002$, the precise redshift at which Frayer et al.\ (1999) detected
CO$(3$--$2)$ emission.  The Ly$\alpha$ emission line is thus
blueshifted by $\sim 400$\,km\,s$^{-1}$ compared to the CO and
H$\alpha$ emission, the size and sense of shift expected for a UV
line affected by outflow absorption. The optical absorption lines are
affected by a similar shift. The flux density in the H$\alpha$ line is
$(5.6 \pm 0.8) \times 10^{-19}$\,W\,m$^{-2}$; that of the [N\,{\sc
ii}] line, which is seen as a shoulder on the H$\alpha$ profile, is
$(1.9 \pm 0.8) \times 10^{-19}$\,W\,m$^{-2}$. The
H$\alpha$/(H$\alpha$+[N\,{\sc ii}]) ratio (0.86) is consistent with
values in normal galaxies rather than those dominated by AGN
(Kennicutt 1983).
 
%
% FIGURE 5
%
\begin{figure}
\centerline{\psfig{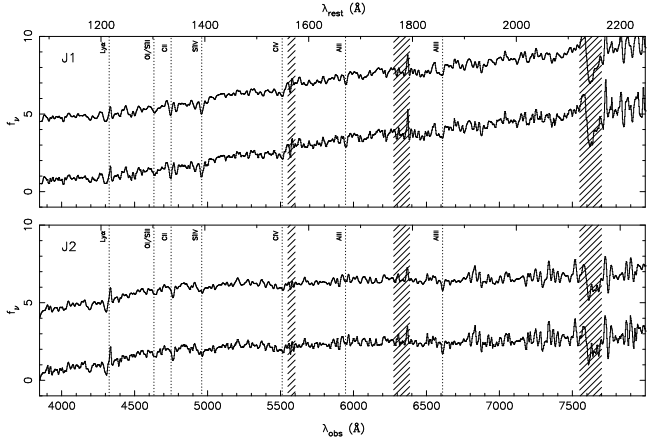}}
\caption{Optical spectra of both SMM\,J14011 J1 and J2 from the LRIS
spectrograph on Keck-{\sc ii}. The lower spectrum in each panel shows
the raw data, while the upper spectrum is smoothed to the instrumental
resolution.  The spectrum of J1 shows a single, narrow, weak Ly$\alpha$
emission line at $z=2.562$, superimposed on a blue continuum. The
spectrum of the fainter component, J2, also shows narrow Ly$\alpha$ at
a similar redshift to that of J1. A number of strong absorption lines
are visible in the continuum of both objects.  The hatched regions are
strongly affected by atmospheric absorption or emission. The optical
lines listed in Table~3 are indicated. The observed wavelength is shown
on the bottom axis; the rest-frame wavelength at $z=2.56$ is shown on
the top axis. The spectra of J1 and J2 are flux calibrated and are
plotted in units of $10^{-29}$\,erg\,cm$^{-2}$\,s$^{-1}$\,Hz$^{-1}$.  }
\end{figure}

%
% FIGURE 6
%
\begin{figure}
\centerline{\hspace{0.3in} \psfig{file=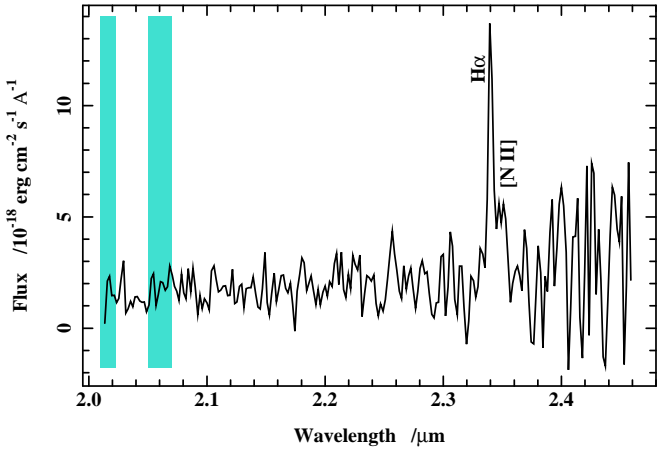,angle=0,width=3.0in}}
\caption{Near-IR spectrum of SMM\,J14011 J1/J2 from the CGS4
spectrograph on UKIRT. We see a clear, narrow line at 2.340$\mu$m
which we identify as Balmer H$\alpha$ at $z=2.565$; there is a
pedestal contribution on the red side of the line from the [N\,{\sc
ii}] line at 6583\AA, but there is no evidence of a broad component to
the H$\alpha$ line as might have been expected if a significant
fraction of the emission was powered by an obscured AGN.}
\end{figure}

\setcounter{table}{2}
%
% Table 3
%
\begin{table}
\hspace*{-0.5cm}\begin{center}
\caption{\hfil Spectral line identifications$^1$ in SMM\,J14011. \hfil }
\begin{tabular}{lcccl}
\noalign{\medskip}
\noalign{\smallskip}
{Line} & {$\lambda_{\rm obs}$}& {$\lambda_{0}$} & {$\>\>z\>\>$} & 
 {Comments} \cr
& (\AA) & (\AA) &  & \cr
\noalign{\medskip}
{\bf J1} & & &  \cr
Ly\,$\alpha$ & 4330.3 & 1215.7 & 2.562  &  Narrow em.\ line\cr
Si\,{\sc ii}/O\,{\sc i} & 4636.1 & 1302/1304 & 2.558 & Abs.\ blend\cr
C\,{\sc ii}  & 4760.2 & 1334.5 & 2.567 & Abs.\ line\cr
Si\,{\sc iv} & 4962.0 & 1394/1403 & 2.552  &  Abs.\ blend\cr 
C\,{\sc iv}  & $\sim$5510 & 1549.0 & 2.558 & Broad abs.\ line\cr
H\,$\alpha$  & 23400 & 6563 & 2.565 & Narrow em.\ line\cr
\noalign{\medskip}
{\bf J2} & & &  \cr
Ly$\alpha$ & 4330.3 & 1215.7 & 2.562  &  Narrow em.\ line,\cr
           &        &        &        & abs.\ 40\AA\ to blue\cr
Si\,{\sc ii}/O\,{\sc i} & 4633.5 & 1302/1304 & 2.556 & Abs.\ blend\cr
C\,{\sc ii}  & 4745.5 & 1334.5 & 2.556 & Abs.\ line\cr
Si\,{\sc iv} & 4958.0 & 1394/1403 & 2.549  & Abs.\ line\cr 
C\,{\sc iv}  & 5508.2 & 1549.0 & 2.556 & Abs.\ line,\cr
             &        &        &       & em.\ line in\cr
             &        &        &       & abs.\ trough?\cr
\noalign{\smallskip}
\end{tabular}
\end{center}

\noindent
$^1$ $\lambda_{\rm obs}$ is the observed wavelength; $\lambda_{0}$ is the
emitted wavelength.

\end{table}

\subsection{SMM\,J14009+0252 (J5)}

%
% FIGURE 7
%
\begin{figure}
\centerline{\psfig{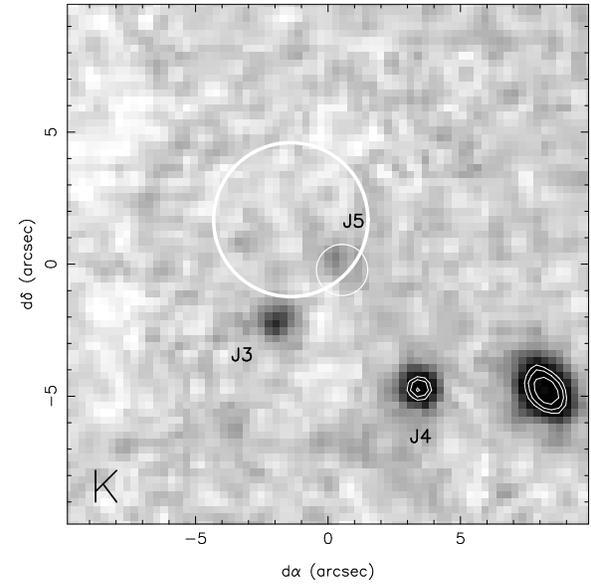}}
\caption{$K$ image of SMM\,J14009, with the possible near-IR counterparts
identified. The large circle shows the positional uncertainty for
the submm source. The small circle marks the radio counterpart; its
1$''$ radius accounts for the combined radio and IR positional
uncertainty. The very faint source J5 lies 0.7$''$ from the position
of the radio counterpart which we associate with the submm emission and
hence we identify this as the likely source of both the radio and submm
emission. The panel is $20''$ square; north is up, east to the left. The
image has been smoothed with a $0.45''$ FWHM Gaussian.}
\end{figure}

A bright 1.4-GHz radio source is identified 2.3$''$ from the nominal
submm position of SMM\,J14009. Owing to the low surface density of
radio sources at these flux limits it is unlikely that this is a
chance superposition and so we identify the radio emission as arising
from the same source as the submm emission.  At 1.4-GHz the radio
emission from SMM\,J14009 is unresolved and relatively strong, with a
peak surface brightness of $529 \pm 16$\,$\mu$Jy\,beam$^{-1}$;
deconvolution with the sythesized beam suggests that its intrinsic
FWHM cannot exceed $1.4''$.

There is no obvious optical counterpart to SMM\,J14009 at the radio
source position in the ground-based imaging, down to apparent
magnitudes of $I>23.0$ or $B>25.0$ (3$\sigma$). This identification of
the source as a `blank' optical field based on the accurate position of
the new radio counterpart differs from that given by Smail et
al.\ (1998a) who, working from the low-resolution submm maps,
identified a `possible' faint counterpart (J3) to SMM\,J14009 on the
ground-based $I$-band image with a 15 per cent likelihood of this being
a chance coincidence\footnote{Another galaxy, J4, still further from
the submm position, was labelled in finders for the spectroscopic
survey of Barger et al.\ (1999). In $I$, J4 is at $14^{\rm h} 00^{\rm
m} 57.^{\rm s}75$, Dec.\ $+02^{\circ} 52' 47.''2$ (J2000); in $3''$
apertures, it has magnitudes of $U=24.14\pm 0.32$, $B=24.93\pm 0.38$,
$I=23.02\pm 0.44$ and $K=19.19\pm 0.05$.}. However, using the 1.4-GHz
map (\S2.1) we are able to rule out the proposed optical counterparts
and instead categorise the source as an optically blank field.  

%
% FIGURE 8
%
\begin{figure*}
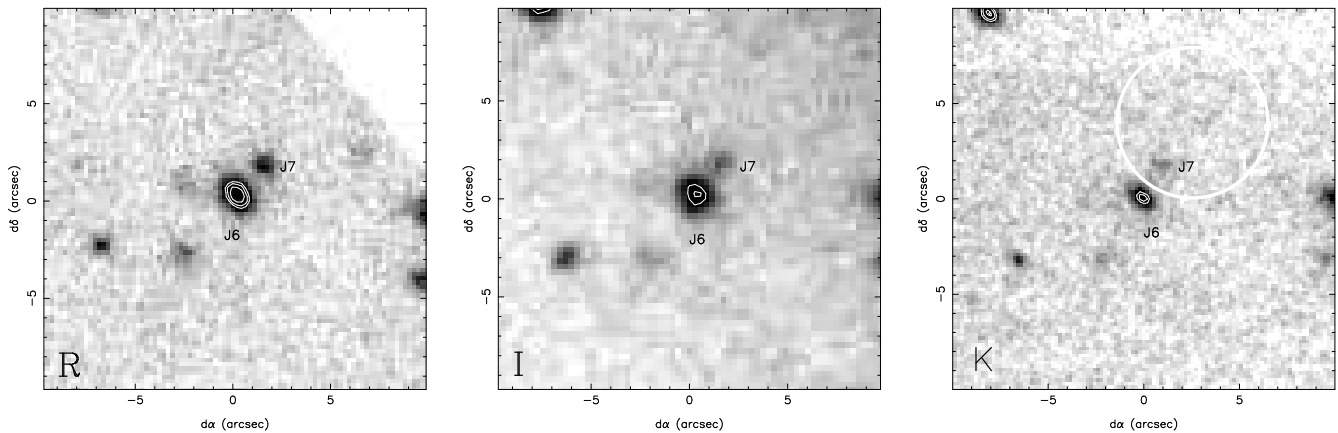

\centerline{
\psfig{file=fig8a.eps,angle=270,width=2.2in} \hspace*{0.1in}
\psfig{file=fig8b.eps,angle=270,width=2.2in} \hspace*{0.1in}
\psfig{file=fig8c.eps,angle=270,width=2.2in}}
\caption{$RIK$ images of SMM\,J14010+0253, with the counterparts, J6
and J7, identified.  Notice the extension of J6 in all passbands and
the very compact nature of J7 in the best seeing $R$ and $K$ images.
In the $K$ image, the circle shows the positional uncertainty for the
submm source. The panels are $20''$ square; north is $5.4^{\circ}$
clockwise in all panels.}
\end{figure*}

In our deep $K$ image of SMM\,J14009 (Fig.~7), the galaxy identified
optically by Smail et al.\ (1998a), J3, is visible ($K=20.33\pm 0.11$
in a $2''$ aperture) $3.3''$ from the position of the 1.4-GHz source
and $2.9''$ from the nominal position of the 850-$\mu$m source.  The
excellent positional accuracy afforded by the radio map makes such
large positional offsets unlikely, and so we rule out J3 as the
origin of the radio and submm emission (although see later).  A faint
source, J5, is visible in the $K$ image but not in the
optical images (Fig.~7, Table~2).  J5 is a far more likely counterpart
for SMM\,J14009, being only $0.7''$ and $1.8''$ from the nominal
positions of the 1.4-GHz and 850-$\mu$m sources respectively ($3.3''$
from J3 at PA 145$^\circ$).

While we do not believe J3 is the source of the submm emission, we do
not rule out a connection between J3 and J5. Indeed, such a relationship
might be expected if an interaction between J3 and J5 triggered the
submm emission seen in the latter.  Such a hypothesis is a natural
consequence of the statistics of merging and interacting systems in the
submm population seen by Smail et al.\ (1998a).  Given the faintness
of J5, spectroscopic observations will be difficult, and so studying a
less-obscured possible companion (J3) may provide valuable information
on the redshift of J5.  Current spectroscopic limits on J3 suggest it
lies somewhere in the redshift range, $z\sim 1.5$--$2.5$ (Barger et
al.\ 1999b).

To try to constrain the likely redshift of J5 we turn to the
long-wavelength spectral properties of the source.  The radio--submm
spectral index, $\alpha^{850}_{1.4}$, (Carilli \& Yun 1999; Blain
1999) is $\alpha^{850}_{1.4}=0.60\pm 0.03$ and indicates a likely
redshift range of $0.7\ls z\ls 2.3$ (Smail et al.\ 2000), matching the 
current limits on the redshift of J3.
However, any contribution in the radio from an AGN would significantly
raise the upper bound (see \S4.2.3).  The 450- to 850-$\mu$m flux
ratio has also been used to make coarse redshift estimates (Hughes et
al.\ 1998): J5 has a ratio of $2.1\pm 0.9$ and this suggests $z\gs
2.8$.  The faint near-IR magnitude of J5 also favours a high redshift unless
the source is {\it extremely} obscured.  Comparing this galaxy to the
well-studied ERO, HR\,10 at $z=1.44$ (Graham \& Dey 1996; Cimatti et al.\ 
1998; Dey et al.\ 1999), we see that J5 is three times more luminous in 
the submm, yet is more than 10 times fainter in $K$. Placing J5 at the same 
low redshift as HR10 would require that the galaxy suffered substantially 
more extinction at rest-frame wavelengths of $\sim 1$\,$\mu$m than HR10, 
which is already a highly-obscured and very red galaxy.  For this reason, we
believe that this source is probably more distant than HR\,10 and that
the low $\alpha^{850}_{1.4}$ estimate may reflect some contribution in
the radio from an AGN.  Finally, comparing the SED of SMM\,J14009 (see
\S4) with other submm-bright galaxies suggests that $3\ls z\ls 5$.

The $K$-corrected distance modulus at $z=4$ for is 48.65 for a
spectral index similar to those of SMM\,J02399 and SMM\,J14011, so J5
would have a $K$-corrected absolute magnitude of $M_K = -26.9$, only
$1.4^{\rm mag}$ fainter than that of J1 and quite plausible given the
large range of UV--IR colours exhibited by the submm galaxy population
(Smail et al.\ 1999).

\subsection{SMM\,J14010+0253 (J6/J7)}

The position of SMM\,J14010+0253 is devoid of radio emission in our
1.4-GHz map, and so we must rely on the submm position to identify the
optical counterpart. The absence of radio emission allows us
to constrain the source to lie at $2<z<4$ from the observed radio--submm
spectral index of $\alpha^{850}_{1.4}>0.82$.  Inspecting Fig.~8, there
is an obvious candidate counterpart in the compact galaxy J6 ($0.88''$
FWHM) along with a fainter companion system, J7, $\sim2''$ away.

To test the identification of J6/J7 as the submm source we obtained
optical and near-IR spectra. Our optical spectrum shows several
features, including redshifted C\,{\sc iv}] 1549\AA\ with a P Cygni
absorption wing, and gives a redshift of $z=2.22\pm 0.02$ for the
source (Fig.~9). While J6 dominates the optical continuum emission
(Fig.~8), the emission line is extended along the slit suggesting that
J7 is at the same redshift as J6.

The near-IR spectrum appears to show relatively weak line features in 
the $K$-band window at 2.088 and 2.113\,$\mu$m (labelled A and B in Fig.~9). 
These features appear in both the positive and negative versions
of the spectrum and are thus likely to be real. They are not offset
significantly along the spatial axis. There are several plausible
identifications for the lines, including [O\,{\sc iii}] 4959, 5007\AA\
at $z \sim 3.22$, but there is no sign of the [O\,{\sc ii}] 3727\AA\ line
in our deep $H$-band spectrum (nor of any other lines) which argues
against $z=3.22$. The redshift obtained from the optical spectrum
suggests that at least one of these lines is likely to be Balmer
H$\alpha$; indeed, the 2.113-$\mu$m feature is consistent with Balmer
H$\alpha$ at the optical redshift, $z = 2.22$.  This would imply a
rest-frame wavelength of $\sim$6484\AA\ for the 2.088-$\mu$m line if
it is associated with the same object, or it could be another
star-forming galaxy close to the line of sight at $z=2.18$.

\subsection{SMM\,J14010+0252 (J8)}

With the exception of the central cluster galaxy discussed
by Edge et al.\ (1999), SMM\,J14010+0252 is the brightest 1.4-GHz 
source in the Abell\,1835 region, with a flux density of 
$1.65\pm 0.03$\,mJy. This radio emission is unresolved in our VLA map. 
Taking this flux, we estimate a radio--submm spectral index of 
$0.2\pm 0.1$ for SMM\,J14010+0252, corresponding to a predicted redshift 
range of $0.0<z<1.0$ (Carilli \& Yun 1999).

Fig.~2 shows two possible bright counterparts for the weak submm
emission: a bright red, early-type galaxy to the north (J8, which is closer to
the radio position, and because of its colours is possibly a cluster member 
or a low-redshift field galaxy at $z\ls 0.5$), with a faint companion in its
halo; and a fainter, redder galaxy to the south which appears to be an
edge-on S0--Sa galaxy in our high-resolution $K$-band image.  We
suggest that J8 is the likely source of the submm and radio emission
and further suggest on the basis of both its colours and its radio--submm
spectral index that this is most likely to be a cluster galaxy.

\section{Discussion}

Abell\,1835 was included in our SCUBA survey because of its extreme
X-ray luminosity and the presence of strongly lensed features in deep
ground-based images (Smail et al.\ 1998b).  These lensed features have
enabled the construction of a mass model of the core of the cluster
(see Blain et al.\ 1999a), which can be used to robustly correct the fluxes
of any background sources for lens amplification.  Until this point,
no quoted observational quantities (e.g.\ Table~2) have been corrected
for the effects of gravitational lensing; from here, however, we will
correct all derived physical quantities, e.g.\ $M_R$, for amplification
by the cluster lens.

%
% FIGURE 9
%
\begin{figure}
\psfig{file=fig9a.eps,angle=270,width=3.35in}

\hspace{0.3in}\psfig{file=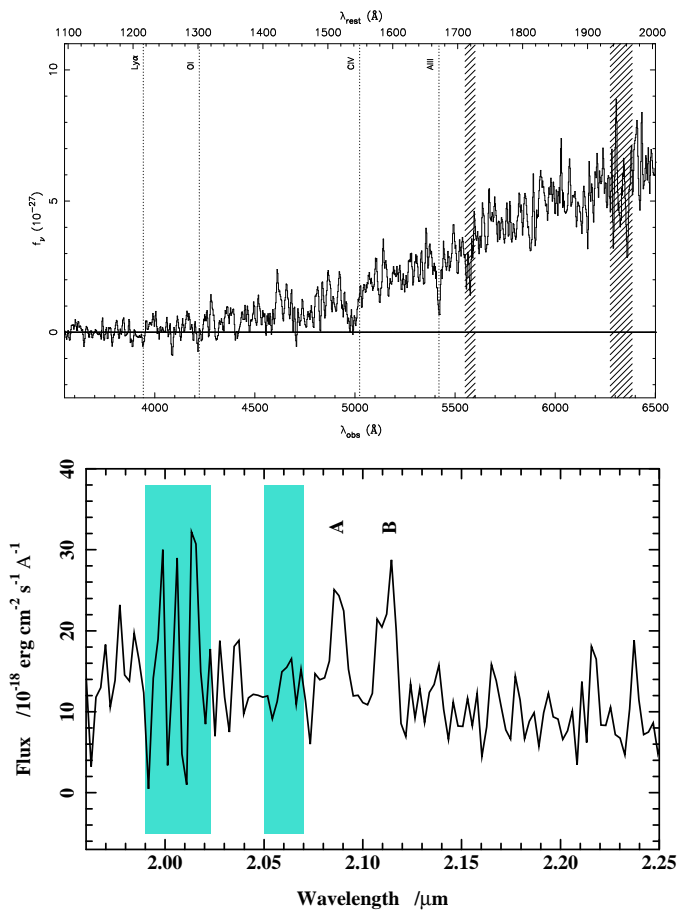,angle=0,width=3.0in}
\caption{Spectrum of SMM\,J14010+0253 J6/J7 from LRIS on Keck-{\sc ii}
({\em top}) and CGS4 on UKIRT ({\em bottom}). In the optical, where
the flux is shown in units of erg\,cm$^{-2}$\,s$^{-1}$\,Hz$^{-1}$, we
see Al\,{\sc ii} in absorption and C\,{\sc iv} blue-shifted in absorption. 
The wavelengths of Ly$\alpha$ and O\,{\sc i} are also marked, 
although the lines are not detected. In the near-IR we see weak line
emission at 2.088 and 2.113\,$\mu$m (marked A and B); B is
presumably Balmer H$\alpha$ at $z=2.2196$. The origin of A is
unknown; it may be a line-of-sight star-forming galaxy at $z=2.18$.}
\end{figure}

Using our lens model we estimate amplifications ($A$, in flux) of the
various sources as: SMM\,J14011 at $z=2.56$, $A=3.0 \pm 0.6$;
SMM\,J14009 assuming $z = 1.0$--$4.0$, $A=1.5\pm 0.2$;
SMM\,J14010+0253 at $z=2.22$, $A= 4.8\pm 2.8$, where the upper bound
is set by the lack of an optical counter-image; and no amplification
of SMM\,J14010+0252, assuming it lies in the cluster. These
amplifications are predominantly due to tangential shearing of the
images, around the centre of the lens at the cD galaxy, and so the spatial 
resolution of the source is finer in the tangential direction as compared 
with that in the radial direction.  For example, for SMM\,J14011 at 
$z\sim 2.5$, an angular scale of $1''$ corresponds to 7.7\,kpc in the 
radial direction, but to only 2.6\,kpc in the tangential direction.

De-amplifying the submm sources, we estimate intrinsic 850-$\mu$m
fluxes of 4.9, 10.4, 0.9 and 4.2\,mJy for SMM\,J14011, SMM\,J14009,
SMM\,J14010+0253 and SMM\,J14010+0252, respectively.  Note that
SMM\,J14010+0253 is therefore one of the faintest submm sources
detected, and the only known representative of the mJy-level submm
population which dominates the submm background (Blain et al.\ 1999a).

\subsection{Morphologies}

The optical and near-IR morphologies of the counterparts to the submm
sources appear compact, but not unresolved, with seeing-corrected FWHM
in the 0.5--1.0$''$ range, corresponding to scales of 2--4\,kpc taking
the lens amplification into account.  There is therefore good evidence
for the submm emission coming from relatively large galaxies with
extended rest-frame optical emission, galaxies not dominated by a
central point source in the rest-frame UV/optical.

The morphological feature linking two of the sources, SMM\,J14011 and
SMM\,J14010+0253, is the presence of pairs of optical counterparts,
separated by 2$''$ ($\sim 5$\,kpc in the source plane).  In both cases
the available spectroscopy suggests that these pairs are physically
associated.  The fainter component of the pairs have rest-frame
$I$-band luminosities about 20--30 per cent of the primary and may thus
represent dynamically important components within the systems,
although it is important to note that that in the optical/near-IR we
do not appear to see mergers of equal luminosity (mass?) components.
These two systems are strikingly similar to the brightest submm source
in our survey, SMM\,J02399 at $z=2.8$ (I98), which also comprises two 
components, separated by 9\,kpc.  The morphological
similarity even stretches to the presence of faint, extended emission
`plumes' around all three pairs, perhaps associated with line-emitting
nebulae.  This provides further support for a large fraction of interacting
systems in the distant submm population (Smail et al.\ 1998a).  The
similarly large proportion of interactions and mergers seen in the
local ULIRG population (Sanders \& Mirabel 1996) argues for the same
physical mechanism, gravitational interaction, triggering the activity
in both the distant submm population and local ULIRGs.

The comparable separations of all three pairs suggests that these
merging systems have very great far-IR luminosities when the components
are relatively widely separated, with projected distances of
5--10\,kpc.  This contrasts with the situation in local ULIRGs, which
are typically most luminous when the two merging components are only 1--2\,kpc 
apart, with a tail of sources with wider separations out to $\gs
10$\,kpc (Murphy et al.\ 1996).  To increase the luminosity of the
system at an earlier stage of the merger, more gas must be moved into
the central regions of the galaxies through bar instabilities (Mihos
\& Hernquist 1996; Bekki, Shioya \& Tanaka 1999).  Massive stellar
bulges in galaxies suppress bar formation in their disks, and so our
results may suggest that while optically luminous, these submm-selected 
merging galaxies do not possess large bulges, in contrast to the
progenitors of local ULIRGs.

\subsection{Probes of buried AGN}

\subsubsection{X-ray emission}

An obvious test of the presence of an AGN in the submm sources is to
search for hard X-ray emission.  We find no evidence of soft X-ray
emission at the level of $S(0.1$--2\,keV$)\ls 10^{-13}$ erg\,s$^{-1}$
(Table~2) from any of the distant submm sources in this field in the
shallow {\it ROSAT} 0.1--2\,keV X-ray image.  While this allows us to
rule out the presence of relatively unobscured AGN in these sources
(Gunn et al.\ 2000, in prep), it will require forthcoming hard X-ray
missions such as {\em Chandra} and {\em XMM} to place stronger limits
on the presence of highly-obscured AGN in these galaxies (Almaini,
Lawrence \& Boyle 1999; Gunn \& Shanks 1999). The sensitivity of {\em
XMM} to hard X-ray emission will be extremely useful in this regard;
however, the observational key may prove to be the higher spatial
resolution of {\em Chandra}, which will be essential to remove background
emission from the extended X-ray-bright cluster.

\subsubsection{Optical/near-IR spectra}

It is remarkable that galaxies such as SMM\,J14011 and SMM\,J02399,
with similar morphologies and broad-band spectral properties and gas
masses suggestive of proto-ellipticals (Frayer et al.\ 1998, 1999),
should have such different emission- and absorption-line
characteristics.

Where the spectrum of SMM\,J02399 shows high-excitation emission lines
typical of a type-{\sc ii} AGN, the optical and near-IR spectra
presented in \S3 for SMM\,J14011 (J1/J2) and SMM\,J14010+0253 (J6/J7)
provide no evidence that their far-IR luminosities are powered by
AGN. For J1/J2 we see the typical rest-frame UV spectrum of a
starburst -- absorption features and a narrow Ly$\alpha$ emission
line.  In the rest-frame optical, the H$\alpha$/(H$\alpha$+[N\,{\sc
ii}]) ratio is consistent with values observed in normal galaxies, and
the H$\alpha$ profile shows no sign of an underlying broad-line
component (often proposed as a probe of buried AGN, although see
\S4.4). For J6/J7, the optical/IR spectra are also reminiscent of a
starburst and we see several absorption lines common to SMM\,J14011
(J1/J2), e.g.\ Al\,{\sc ii} and C\,{\sc iv}.

\subsubsection{Mid-IR spectra}

The mid-IR fluxes expected for SMM\,J14009 and SMM\,14011 (Fig.~10)
are around 1---10\,mJy at 25\,$\mu$m. Detections at this level are
achievable with Michelle, the mid-IR imaging spectrograph soon to be
delivererd to UKIRT. Future observations of these sources with
Michelle, on UKIRT and Gemini, will determine the amount of warm dust
emission from these galaxies, another possible diagnostic/probe of the
presence of an AGN power source.

\subsubsection{Radio emission}

The radio properties of the submm-selected galaxy population have been
discussed by Smail et al.\ (2000). Half of the fifteen lensed
submm-selected galaxies in the sample used by Smail et al.\ were
detected at 1.4\,GHz, and of those, only three were bright enough to
have been identified by wide-field radio surveys such as NVSS and
FIRST (Condon et al.\ 1998; Becker, White \& Helfand 1995) ---
SMM\,J02399$-$0134, SMM\,J02399 and SMM\,J14009. The remaining
detections were at the $\ls 100$\,$\mu$Jy level, and are thus
consistent with the absence of radio-loud AGN.

At $z\sim 4$, the 1.4-GHz flux from SMM\,J14009 would give a
rest-frame power of $6.5 \times 10^{23}$\,W\,Hz$^{-1}$\,sr$^{-1}$ at
7\,GHz, greater than that of SMM\,J02399, an order of magnitude
greater than that of SMM\,J14011, and unheard of amongst low-redshift
star-forming galaxies (Mobasher et al.\ 1999). However, it is still
more than an order of magnitude less than that of known $z\sim 4$
radio galaxies (e.g.\ van Breugel et al.\ 1999). If SMM\,J14009 lies
at $z<4$ then the requirement for AGN-powered radio emission to
explain the SED is eased, and for $0.7 < z < 2.3$ the radio emission
becomes consistent with a pure starburst. However, we have already
argued that such a low redshift is unlikely (\S3.2). We conclude that
the radio emission from SMM\,J14009 includes a contribution from an
AGN.

For SMM\,J02399 -- an interaction/merger in which one component is
known to harbour an AGN from its optical spectrum -- the rest-frame
power at 5.3-GHz is $2.6 \times 10^{23}$\,W\,Hz$^{-1}$\,sr$^{-1}$
(Ivison et al.\ 1998), which is extremely rare amongst star-forming
galaxies, although the far-IR luminosity of SMM\,J02399 is phenomenal
($\gs 10^{13}$\,L$_{\odot}$), and so one would therefore expect even
its purely starburst-driven radio luminosity to be similarly extreme.
Indeed, the observed 1.4--8.5\,GHz spectral index is in agreement with
optically thin synchrotron emission (Ivison et al.\ 1999), and the
radio--submm spectral index is consistent with the spectroscopic
redshift ($z=2.80$) based on the usual far-IR/radio correlation for
starbursts (Carilli \& Yun 1999). The 1.4-GHz morphology is also
consistent with contributions from both of the optical/near-IR
components (Owen et al., in prep). It is not possible to rule out an
AGN contribution to the radio power of this galaxy, although it seems
unlikely that an AGN provides the sole contribution.

The only other submm-selected galaxy where the presence of an AGN
cannot be disputed is SMM\,J02399$-$0134, a Seyfert-2 ring galaxy
(Soucail et al.\ 1999). However, its radio luminosity and 
radio--submm spectral index are both consistent with a pure starburst 
at the spectroscopic redshift ($z=1.06$), and so the radio emission 
is not a useful AGN diagnostic in this case.

The 1.4-GHz flux from SMM\,J14011 translates into a rest-frame power
at 5.0\,GHz of $4.1 \times 10^{22}$\,W\,Hz$^{-1}$\,sr$^{-1}$ which is
reasonable, if rare, for a star-forming galaxy (Mobasher et al.\
1999). The radio--submm spectral index is consistent with the known
redshift, and again there is no evidence to support an AGN
contribution.

Interestingly, the lesson from the radio seems to be that 1.4-GHz
observations are not necessarily a good test of the presence of an
AGN in submm-selected galaxies, though they are clearly very helpful
when pinpointing counterparts and constraining redshifts. Even for
SMM\,J14009, the presence of a buried AGN cannot be inferred from the
radio data alone, at least not at $\sim 5''$ resolution.

In summary, the existing {\em ROSAT} X-ray imaging and optical/near-IR
spectroscopy of sources in this field indicate that they do not
contain unobscured AGN. However, it is difficult to place limits on the 
presence of obscured AGN at present, in the absence of mid-IR and hard 
X-ray images.

%
% FIGURE 10
%
\begin{figure*}
\centerline{\psfig{file=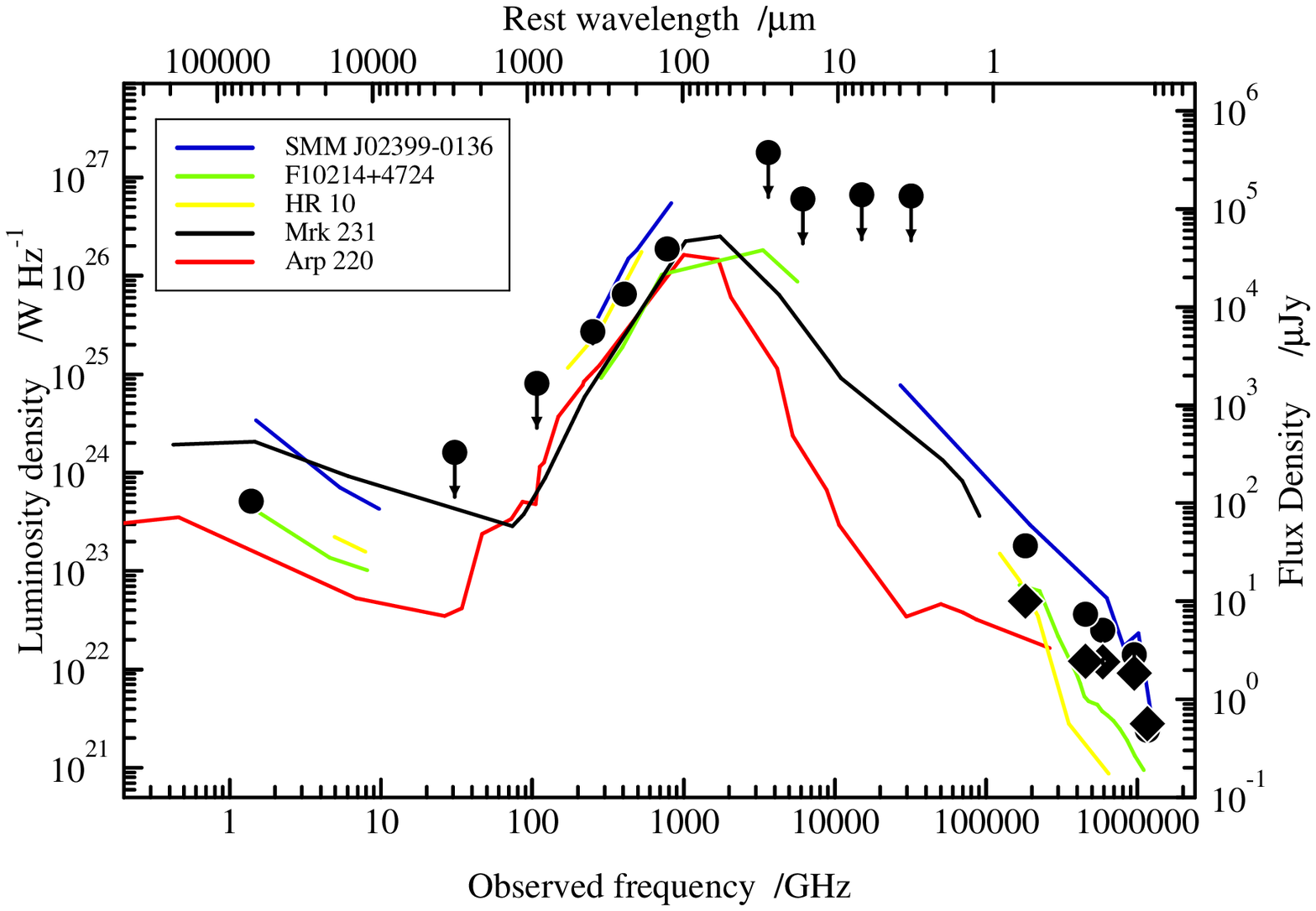,width=3in}
\hspace*{0.5in} \psfig{file=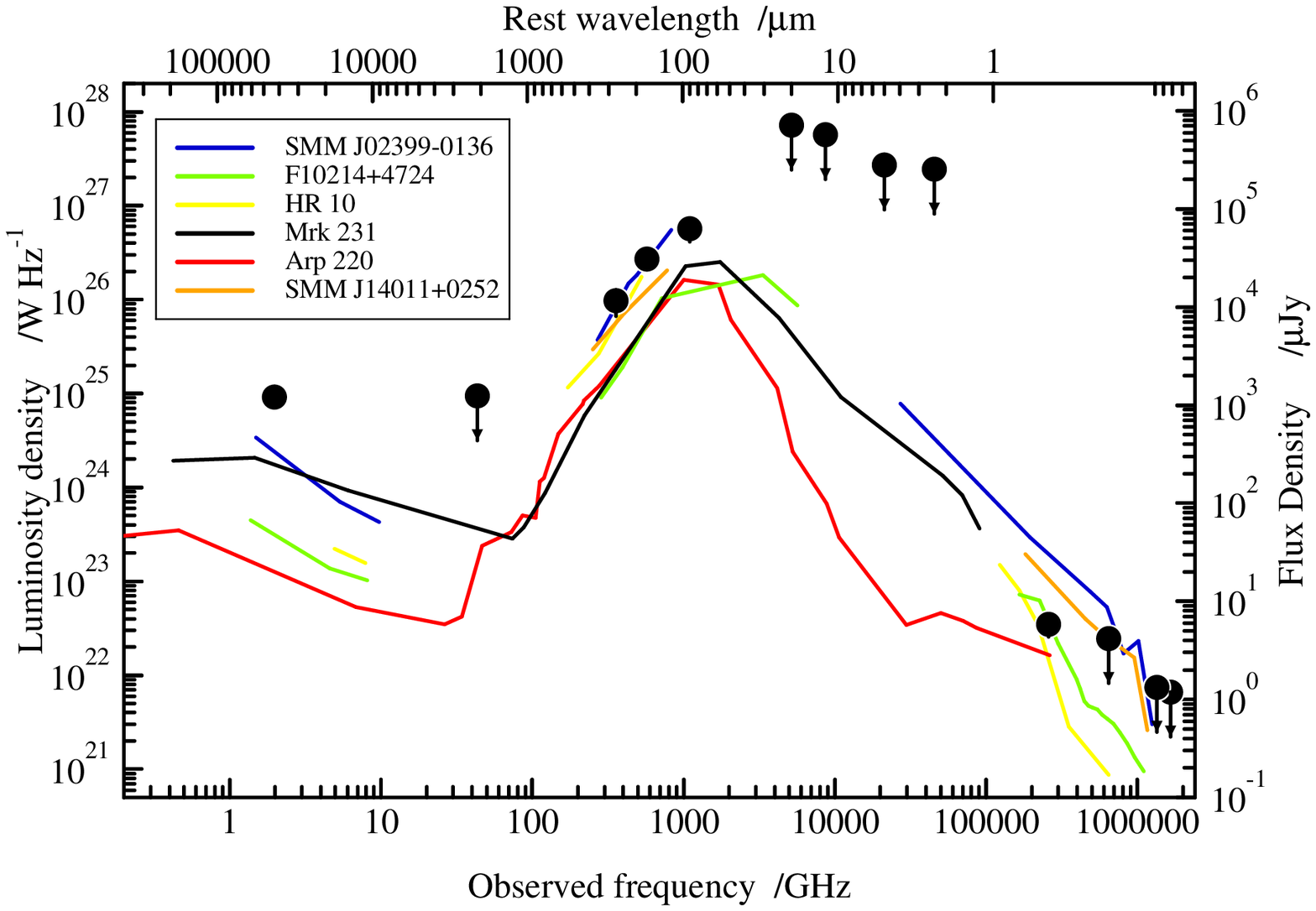,width=3in}}
\caption{{\bf a) (left)} The SED of SMM\,J14011 J1 between the
rest-frame radio and UV wavebands, represented by filled circles. J2
is represented by filled diamonds. Beyond 1\,$\mu$m (rest frame) the
points represent J1 + J2. The right-hand scale gives the flux density
for SMM\,J14011.  ~{\bf b) (right)} The SED of SMM\,J14009 J5 between
the rest-frame radio and UV wavebands. The right-hand scale gives the
observed flux densities for SMM\,J14009. The scales on the top and
left assume $z=4$ for SMM\,J14009 and are corrected for 0.44$^{\rm
mag}$ of amplification.  In both plots we show for comparison the SEDs
of other well-known submm- and IR-selected galaxies, SMM\,J02399
(I98), {\em IRAS} F10214+4724 (Rowan-Robinson et al.\ 1993; Barvainis
et al.\ 1995), Arp~220, Mrk~231 (Hughes, Davies \& Ward, in prep) and
HR~10 (Dey et al.\ 1999). The lines are broken in regions where only
upper limits to the SEDs are available. {\em IRAS} F10214+4724,
SMM\,J02399 and SMM\,J14011 are corrected for lensing by factors of
30, 2.4 and 3.0, respectively (Broadhurst \& Leh\'ar 1995; I98).}
\end{figure*}

\subsection{SEDs}

In Fig.~10 we plot the rest-frame UV---radio SED of SMM\,J14011 and
SMM\,J14009 (assuming $z=4$), deduced from our observations and
corrected for amplification by the cluster lens.  For SMM\,J14011 at
rest-frame wavelengths longer than about 1\,$\mu$m, this is
effectively the SED of both J1 and J2. We have also included upper
limits from the {\it IRAS} All-Sky Survey at 12, 25, 60 and
100\,$\mu$m (Table~2).  The SEDs are poorly constrained at wavelengths
between about 1 and 100\,$\mu$m. At the longer wavelengths observed by
SCUBA, we find $\alpha = 1.90^{+0.85}_{-0.75}$ for SMM\,J14011 across
the rest-frame wavelength range 130--380\,$\mu$m, marginally
consistent with the characteristic spectral index, $\alpha \simeq 3\pm
1$, of optically-thin emission from dust grains.  For SMM\,J14009, the
observed far-IR SED is a little steeper, with $\alpha =
2.2^{+1.1}_{-0.8}$ between 1350 and 850\,$\mu$m, flattening to $\alpha
= 1.2^{+0.6}_{-0.7}$ between 850 and 450\,$\mu$m.

Simple fits to these SEDs cannot, of course, constrain the dust
temperature, although the data for both sources are consistent with
the $T_{\rm d}$ range of 35--50\,{\sc k} found for other dusty,
high-redshift galaxies (I98; Benford et al.\ 1999; see also Dunne et
al.\ 2000). Higher temperatures can be fitted if the opacity of the
dust becomes significant at wavelengths of about 100--200\,$\mu$m.
Adopting $k_{\rm d} = 0.15 [\lambda_0/800\,\mu{\rm m}]^{-1.5}$ for the
standard dust emission parameter in order to facilitate comparisons
with other distant objects, where $\lambda_0$ is the rest-frame
wavelength, we estimate a dust mass of $M_{\rm d}= 2.3^{+1.1}_{-0.6}
\times 10^8$\,M$_{\odot}$ for SMM\,J14011 if $T_{\rm d} = 50\pm
10$\,{\sc k}.  For SMM\,J14009, taking $T_{\rm d} = 50$\,{\sc k}, the
dust masses implied by the submm emission for $z=3 \rightarrow 5$
would be in the range $(3.5 \leftarrow 4.5) \times 10^8$\,M$_{\odot}$,
midway between the values derived using the same assumptions for
SMM\,J14011 and SMM\,J02399.

The luminosity of SMM\,J14011 in the far-IR waveband between observed
wavelengths of 20 and 1000\,$\mu$m, $L_{\rm FIR}$, is about $6 \times
10^{12}$\,L$_{\odot}$, although there remains a large uncertainty due
to the unconstrained mid-IR spectrum.  If we continue to assume that
$z\sim 4$, then the luminosity of SMM\,J14009 in the far-IR waveband,
$L_{\rm FIR}$, will be {\it at least} $10^{13}$\,L$_{\odot}$.
SMM\,J14010+0253, with an intrinsic submm flux of about 1\,mJy, has a
relatively low far-IR luminosity by the standards of other distant
submm-selected galaxies, about $3\times
10^{11}$\,L$_{\odot}$. Therefore while both SMM\,J14009 and
SMM\,J14011 class as high-redshift ultraluminous or hyperluminous IR
galaxies (ULIRG/HLIRG), SMM\,J14010+0253 is one of the first luminous
IR galaxies (LIRG) identified at high redshift.

For SMM\,J14010+0252, assuming it is a cluster member with a close
companion, we derive a dust mass of about $8 \times 10^7$\,M$_{\odot}$
(for 50-{\sc k} dust) and a modest far-IR luminosity of $\approx 5
\times 10^{11}$\,L$_{\odot}$.

The SEDs of several other luminous far-IR galaxies are also shown in
Fig.~10.  Comparing the far-IR luminosities of SMM\,J14011 and
SMM\,J14009 with that of SMM\,J02399 (I98), we see that SMM\,J14011 is
about 25 per cent less luminous, lying somewhere between the
luminosity of SMM\,J02399 and F10214+4724 (Rowan-Robinson et al.\
1993), while at $z=4$ the luminosity of SMM\,J14009 would be
comparable to that of SMM\,J02399.  Overall luminosities aside,
however, the three submm-selected galaxies SMM\,J14011, SMM\,J14009
and SMM\,J02399 have very similar SEDs.  The only significant
difference occurs in the radio waveband. SMM\,J14011 is somewhat less
luminous and SMM\,J14009 is slightly more luminous than SMM\,J02399.
We interpret this sequence as arising from an increasing contribution
from an active nucleus in SMM\,J14011, SMM\,J02399 and SMM\,J14009
respectively. However, apart from a weak contribution in the radio,
this change makes little overall difference to the SEDs of the
sources.

\subsection{SFRs}

We have three independent estimates of SFR in these submm sources:
their luminosities in the far-IR, radio or H$\alpha$ line (we
disregard estimates based on the rest-frame UV flux due to
difficulties of dealing with obscuration in these dusty systems).

If the dust in these sources is heated primarily by young, massive
stars, then $L_{\rm FIR} \sim 10^{12}$L$_\odot$ corresponds to a
formation rate of 650\,M$_{\odot}$\,yr$^{-1}$ adopting a Salpeter
initial mass function (IMF) extending from 0.1 to 100\,M$_{\odot}$
(Thronson \& Telesco 1986).  However, it has been suggested by Zepf \&
Silk (1996) and Bromm, Coppi \& Larson (2000), amongst others, that
the IMF in proto-ellipticals may be top-heavy, and so the same $L_{\rm
FIR} \sim 10^{12}$L$_\odot$ could be produced just from O, B and A
stars with a SFR of about $210$\,M$_{\odot}$\,yr$^{-1}$. In the
absence of a significant non-thermal radiation source, this factor of
three range in the estimated SFRs provides a reasonable indication of
the level of uncertainty in this calculation.  Hence for the far-IR
luminosities of SMM\,J14011 ($6 \times 10^{12}$\,L$_{\odot}$),
SMM\,J14009 ($\sim 10^{13}$\,L$_{\odot}$), SMM\,J14010+0252 ($5 \times
10^{11}$\,L$_{\odot}$) and SMM\,J14010+0253 ($3 \times
10^{11}$\,L$_{\odot}$), we estimate corresponding SFRs in the range
1260--3900, 2100--6500, 105--325 and 60--195\,M$_{\odot}$\,yr$^{-1}$,
respectively. These rates are substantial and if sustained for $\gs
10^8$--10$^9$\,yrs would be capable of forming the bulk of a luminous,
L$^\ast$, galaxy. The energy input into the interstellar medium of the
galaxy from such a burst of star formation would probably be
sufficient to trigger a galactic wind and so the mechanism may be
self-regulating.

The 1.4-GHz radio emission from SMM\,J14011 is consistent with an SFR
of order 650\,M$_{\odot}$\,yr$^{-1}$, for stars $\ge8$\,M$_{\odot}$
(Condon 1992), within a factor of two of our far-IR-based estimate for
high-mass stars. The H$\alpha$ luminosity of SMM\,J14011 (J1/J2),
$L_{{\rm H}\alpha} \simeq 9.1 \times 10^{35}$\,W (after correcting for
lensing) yields a much lower SFR, in the range
35--350\,M$_{\odot}$\,yr$^{-1}$ (Kennicutt 1983; Hill et al.\ 1994;
Gallego et al.\ 1995; Barbaro \& Poggianti 1997). This is also
consistent with the findings of Cram et al.\ (1998), who showed that
H$\alpha$-based estimates appeared to severely underestimate the SFR
in intensely star-forming galaxies.  One plausible explanation is that
there may be significant extinction due to dust even at rest-frame
wavelengths of $\sim$6600\AA. A value of $A_V$ in the range $1.8\ls
A_V\ls 6.5$ would account for the depression of the apparent H$\alpha$
luminosity.  If correct, this suggests that searching for broad
components to H$\alpha$ in these dusty galaxies may not be a
particularly useful probe of buried AGN.

At $z\sim 4$, the radio flux from SMM\,J14009 would indicate a
high-mass SFR more than an order of magnitude higher than that
suggested by the far-IR luminosity. We believe this is probably due to
a dominant AGN contribution to the radio luminosity (\S4.2.3). With no
spectroscopic observations there is no limit on H$\alpha$ for this
galaxy.

The H$\alpha$ luminosity of SMM\,J14010+0253 (J6/J7), taking only the
line at 2.113\,$\mu$m (Fig.~9) is similar to that of J1/J2: $\sim 8
\times 10^{35}$\,W, and so the implied SFR is also similar --
30--300\,M$_{\odot}$\,yr$^{-1}$. This brackets the values deduced from
the far-IR luminosity of SMM\,J14010+0253, leaving little room for
additional reddening at the wavelength of H$\alpha$.

\section{Conclusions}

We have presented the results of observations of three submm-selected 
galaxies seen through the rich cluster lens, Abell\,1835, and two others 
which probably lie within the cluster.

The presence of large amounts of dust in these systems indicates that
substantial past/current star formation has/is taking place, while the
large reservoirs of gas seen in those systems studied in CO (e.g.\
Frayer et al.\ 1999) show that they contain sufficient fuel to
continue star formation at this level for $\sim$1\,Gyr.  In the
process, they would form a mass of stars typical of a luminous galaxy,
10$^{11}$--$10^{12}$\,M$_\odot$.  The apparently high SFRs, high
dust/metal content and large gas masses are all requirements of
massive proto-elliptical galaxies. The identification of the formation
of massive ellipticals in single, intense events at relatively high
redshifts $z>2$ may be difficult to achieve within present
semi-analytic galaxy-formation models (Baugh et al.\ 1998) unless
these systems inhabit high-density regions in the early Universe.
Further study of the environments of distant submm galaxies may be a
fruitful avenue to more fully understand the nature and subsequent
evolution of these galaxies.

We now have detailed information on the presence or absence of AGN in
seven submm-selected galaxies from our survey: SMM\,J02399$-$0136
(type {\sc ii}, I98), SMM\,J02399$-$0134 (type {\sc ii}, Soucail et
al.\ 1999), SMM\,J04433+0210 and SMM\,J09429+4658 (no evidence for or
against an AGN in either of these; Smail et al.\ 1999),
SMM\,J14009+0252 (probable AGN, type unknown), SMM\,J14010+0253
(starburst) and SMM\,J14011+0252 (starburst).

Thus there is evidence of one form or another for AGN contributions to
the far-IR luminosities of three of the seven submm-selected galaxies
for which detailed observations have been acquired. The sample should
not be biased strongly towards AGN by strong emission lines, since we
include all objects for which detailed observational follow-up exists,
not limiting ourselves to those with spectroscopic redshifts. An
observational bias towards the brightest galaxies in the Smail et al.\
(1998a) sample has been largely avoided since the observational
follow-up has followed the seasons and has not concentrated on the
brightest sources. This high rate of AGN activity is roughly
consistent with that seen in similarly luminous local samples of
ULIRGs (Sanders \& Mirabel 1996).

The growing number of pairs of galaxies spectroscopically identified
as counterparts to submm sources supports our earlier suggestion
(Smail et al.\ 1998a) that merging is the mechanism triggering the
activity in the submm galaxy population, also in line with what is
known of local ULIRGs.

Finally, we highlight SMMMJ14011, the first submm source in which
detailed observations have not uncovered an AGN. The inferred SFR is
several 1000\,M$_\odot$\,yr$^{-1}$, and so we are faced with what
appears to be a proto-elliptical galaxy processing $\sim
10^{11}$\,M$_{\odot}$ of molecular gas into stars on a timescale of
less than 1\,Gyr. High-resolution imaging in hard X-rays with {\it
Chandra} (rest-frame 10--40\,keV) should be able to identify any AGN
that is present but currently undetected, even a Compton-thick AGN.
If this source -- an incontravertibly starburst judging by data
obtained in the optical, near-IR and radio wavebands -- contains a
buried AGN, then there is a good chance that a large fraction of the
bright submm galaxy population also contains AGN.

\subsection*{ACKNOWLEDGEMENTS}

UKIRT is operated by the Joint Astronomy Centre (JAC) on behalf of the
United Kingdom Particle Physics and Astronomy Research Council
(PPARC). The JCMT is operated by the JAC on behalf of PPARC, the
Netherlands Organisation for Scientific Research, and the National
Research Council of Canada. NRAO is operated by Associated
Universities Inc., under a cooperative agreement with the National
Science Foundation. RJI, AWB and IRS acknowledge support from PPARC,
the Raymond and Beverly Sackler Foundation and the Royal Society.  We
acknowledge useful conversations and help from Walter Gear, Tom
Geballe, Katherine Gunn, Chris Mihos and Ian Robson.

\clearpage
\addtolength{\textwidth}{1.0cm}
\setcounter{table}{1}
%
% Table 2
%
\begin{table*}
\begin{center}
\caption{The observed properties of the submm sources.}
\begin{tabular}{lccccccl}
\noalign{\medskip}
Property &\multicolumn{2}{c}{SMM\,J14011+0252}& SMM\,J14009+0252 & \multicolumn{2}{c}{SMM\,J14010+0253}  & SMM\,J14010+0252 & Comment \cr
         & J1 & J2                            &  J5 &        J6 & J7 & J8 & \cr
\noalign{\medskip}
$\alpha$(J2000)& \hbox{$14^{\rm h} 01^{\rm m} 04.97^{\rm s}$} & \hbox{$14^{\rm h} 01^{\rm m} 04.85^{\rm s}$} & $14^{\rm h} 00^{\rm m} 57.57^{\rm s}$&
\hbox{$14^{\rm h} 01^{\rm m} 03.09^{\rm s}$} & \hbox{$14^{\rm h} 01^{\rm m} 03.01^{\rm s}$} & \hbox{$14^{\rm h} 01^{\rm m} 00.57^{\rm s}$} &  $I$ band$^1$  \cr
$\delta$(J2000)& \hbox{$+02^{\circ} 52' 24.6''$}& \hbox{$+02^{\circ} 52' 25.3''$} &$+02^{\circ} 52' 49.1''$ & \hbox{$+02^{\circ} 53' 12.0''$}& \hbox{$+02^{\circ} 53' 13.5''$} & \hbox{$+02^{\circ} 51' 50.5''$} & ($\pm 1''$).\cr
\noalign{\medskip}
$\alpha$(J2000)&\multicolumn{2}{c}{$14^{\rm h} 01^{\rm m} 04.96^{\rm s}$}& $14^{\rm h} 00^{\rm m} 57.55^{\rm s}$ & ... & ... & $14^{\rm h} 01^{\rm m} 00.53^{\rm s}$& 1.4\,GHz \cr
$\delta$(J2000)&\multicolumn{2}{c}{$+02^{\circ} 52' 23.5''$}& $+02^{\circ} 52' 48.6''$ & ... & ... & $+02^{\circ} 51' 49.4''$& ($\pm 0.3''$).\cr
\noalign{\medskip}
$\alpha$(J2000)&\multicolumn{2}{c}{$14^{\rm h} 01^{\rm m} 04.96^{\rm s}$}&$14^{\rm h} 00^{\rm m} 57.68^{\rm s}$ & \multicolumn{2}{c}{$14^{\rm h} 01^{\rm m} 02.92^{\rm s}$} & $14^{\rm h} 01^{\rm m} 00.55^{\rm s}$& 850\,$\mu$m \cr
$\delta$(J2000)&\multicolumn{2}{c}{$+02^{\circ} 52' 25.5''$}& $+02^{\circ} 52' 49.9''$ & \multicolumn{2}{c}{$+02^{\circ} 53' 15.9''$}& $+02^{\circ} 51' 45.6''$& ($\pm 3''$)$^2$.\cr
\noalign{\medskip}
Optical $z$&\hbox{$2.559\pm 0.006$}&\hbox{$2.556\pm 0.005$} & ... & \multicolumn{2}{c}{$2.22\pm 0.02$} & ... & \cr
CO $z^3$  &\multicolumn{2}{c}{$2.565\pm 0.002$}& ... &  \multicolumn{2}{c}{...} & ... & \cr
\noalign{\medskip}
Flux at: &&&&&&& \cr
\noalign{\smallskip}
$\>\>12\,\mu$m&\multicolumn{2}{c}{$3\sigma < 145$\,mJy}& $3\sigma<140$\,mJy& \multicolumn{2}{c}{$3\sigma < 120$\,mJy}&$3\sigma < 115$\,mJy&{\em IRAS} \cr
$\>\>25\,\mu$m&\multicolumn{2}{c}{$3\sigma < 150$\,mJy}& $3\sigma<155$\,mJy& \multicolumn{2}{c}{$3\sigma < 155$\,mJy}&$3\sigma < 120$\,mJy&({\sc xscanpi}). \cr
$\>\>60\,\mu$m&\multicolumn{2}{c}{$3\sigma < 135$\,mJy}& $3\sigma<330$\,mJy& \multicolumn{2}{c}{$3\sigma < 145$\,mJy}&$3\sigma < 140$\,mJy& \cr
$\>\>100\,\mu$m&\multicolumn{2}{c}{$3\sigma < 400$\,mJy}& $3\sigma<415$\,mJy& \multicolumn{2}{c}{$3\sigma < 560$\,mJy}& $3\sigma < 490$\,mJy&\cr
\noalign{\smallskip}
$\>\>450\,\mu$m&\multicolumn{2}{c}{$41.9 \pm 6.9$\,mJy}&$32.7 \pm 8.9$\,mJy & \multicolumn{2}{c}{$3\sigma < 18$\,mJy}& $3\sigma < 18$\,mJy&SCUBA$^4$.\cr
$\>\>850\,\mu$m&\multicolumn{2}{c}{$14.6 \pm 1.8$\,mJy}& $15.6 \pm 1.9$\,mJy & \multicolumn{2}{c}{$4.3 \pm 1.7$\,mJy}& $4.2 \pm 1.7$\,mJy& \cr
$\>\>1350\,\mu$m&\multicolumn{2}{c}{$6.06 \pm 1.46$\,mJy}& $5.57\pm 1.72$\,mJy& \multicolumn{2}{c}{...} & ... & \cr
\noalign{\smallskip}
$\>\>3.0$\,mm&\multicolumn{2}{c}{$3\sigma < 1.8$\,mJy}& ... &\multicolumn{2}{c}{...} & ... & OVRO$^3$. \cr
$\>\>1.05$\,cm&\multicolumn{2}{c}{$3\sigma < 0.54$\,mJy}&$3\sigma < 0.54$\,mJy& \multicolumn{2}{c}{$3\sigma < 0.54$\,mJy}& $3\sigma < 0.54$\,mJy& OVRO$^5$. \cr
$\>\>21.5$\,cm&\multicolumn{2}{c}{$115 \pm 30$\,$\mu$Jy}&$529 \pm 30$\,$\mu$Jy& \multicolumn{2}{c}{$3\sigma < 48$\,$\mu$Jy} & $1.65 \pm 0.03$\,mJy& VLA. \cr
\noalign{\smallskip}
$\>\>U_{\rm ap}$&\hbox{$23.82\pm 0.14$}&\hbox{$23.65 \pm 0.12$}&$3\sigma > 24.2$ & \multicolumn{2}{c}{$23.65\pm 0.20$}& $22.33\pm 0.05$& Aperture \cr
$\>\>B_{\rm ap}$&\hbox{$22.83\pm 0.03$}&\hbox{$23.29\pm 0.04$} &$3\sigma > 25.0$ & \multicolumn{2}{c}{$23.49\pm 0.10$}& $21.72\pm 0.03$& mags \cr
$\>\>R_{\rm ap}$&\hbox{$21.77\pm 0.02$}&\hbox{$22.57\pm 0.05$} & ... & \hbox{$21.23\pm 0.02$}&\hbox{$22.90\pm 0.07$}& ... & ($3''$) \cr
$\>\>I_{\rm ap}$&\hbox{$21.10\pm 0.03$}&\hbox{$22.29\pm 0.06$} & $3\sigma > 23.0$ & \hbox{$20.58\pm 0.02$}&\hbox{$22.41\pm 0.16$}& $18.31\pm 0.01$& \cr
$\>\>K_{\rm ap}$&\hbox{$17.97 \pm 0.03$}&\hbox{$19.37 \pm 0.10$}& $21.23\pm 0.33$ & \hbox{$18.32\pm 0.05$}&\hbox{$20.03\pm 0.20$}& $15.31\pm 0.01$& \cr
$\>\>K_{\rm tot}$&\hbox{$17.76 \pm 0.10$}&\hbox{$19.16 \pm 0.14$}&$20.96\pm 0.36$ & \multicolumn{2}{c}{$17.43\pm 0.05$}& $14.65\pm 0.02$ & Total mag.\cr
\noalign{\smallskip}
$\>\>J_{\rm tot}$&\multicolumn{2}{c}{$19.34\pm 0.06$}& ... & \multicolumn{2}{c}{...} & ... & Total mag.\cr
\noalign{\medskip}
$\>\>${0.1--2.0\,keV}&\multicolumn{2}{c}{$3\sigma < 8 \times 10^{-14}$}& $3\sigma < 7.5 \times 10^{-14}$ & \multicolumn{2}{c}{$3\sigma < 16 \times 10^{-14}$}&$3\sigma < 13 \times 10^{-14}$&HRI$^6$. \cr
\end{tabular}
\end{center}

\noindent
$^1$ $K$-band position for J5.\hfill~

\noindent
$^2$ Positional accuracy is $\pm 4''$ for SMM\,J14010+0253 and SMM\,J14010+0252.\hfill~

\noindent
$^3$ Frayer et al.\ (1999).\hfill~

\noindent
$^4$ Errors include a contribution from our uncertainty in the
absolute calibration (10 per cent).\hfill~

\noindent
$^5$ Cooray et al.\ (1998).\hfill~

\noindent
$^6$ erg\,s$^{-1}$\,cm$^{-2}$, 16$''$ aperture, unabsorbed (from the
{\em ROSAT} HEASARC archive at GSFC).\hfill~
\end{table*}

\end{document}